\pdfoutput=1
\documentclass[a4paper]{jpconf}
\usepackage{enumerate}
\usepackage{amssymb}
\usepackage{amsmath}
\usepackage{graphicx}

\newcommand{\be}{\begin{equation}}
\newcommand{\ee}{\end{equation}}
\newcommand{\ba}{\begin{eqnarray}}
\newcommand{\ea}{\end{eqnarray}}

\newcommand{\erfc}{\mathrm{erfc}}

\begin{document}
\title{An analytical model for a full wind turbine wake}

\author{Aidan Keane, Pablo E. Olmos Aguirre, Hannah Ferchland, Peter Clive, Daniel Gallacher}

\address{SgurrEnergy Ltd, 225 Bath Street, Glasgow, G2 4GZ, UK.}

\ead{aidan.keane@sgurrenergy.com}

\begin{abstract}
An analytical wind turbine wake model is proposed to predict the wind velocity distribution
for all distances downwind of a wind turbine, including the near-wake. This wake model
augments the Jensen model and subsequent derivations thereof, and is a direct
generalization of that recently proposed by Bastankhah and Port\'{e}-Agel.
The model is derived by applying conservation of mass and momentum in the context of actuator disk theory,
and assuming a distribution of the double-Gaussian type for the velocity deficit in the wake.
The physical solutions are obtained by appropriate mixing of the waked- and freestream velocity
deficit solutions, reflecting the fact that only a portion of the fluid particles passing
through the rotor disk will interact with a blade.
\end{abstract}

\section{Introduction}
An analytical wind turbine wake model is proposed to predict the wind velocity distribution for
all distances downwind of a wind turbine, including the near-wake. The model is derived by
applying conservation of mass and momentum in the context of actuator disk theory,
and assuming a distribution of the double-Gaussian-type for the velocity deficit in the wake.

In 1983 Jensen \cite{jensen1983} proposed the first analytical model for a wind turbine wake.
This model has become the de facto industry standard, and has been developed further by
Fransden \etal \cite{frandsen2006} and more recently by Bastankhah and
Port\'{e}-Agel \cite{bastankhah2014}. These wake models are based upon mass and momentum conservation,
and either a top-hat distribution \cite{jensen1983}, \cite{frandsen2006}, or a Gaussian distribution
\cite{bastankhah2014} for the velocity deficit.
The Ainslie model \cite{ainslie1988} is based on a numerical solution of the Navier-Stokes equations
and uses a Gaussian near-wake approach.
The model proposed by Larsen \cite{larsen1988}, \cite{larsen1996} is based on the Prandtl turbulent boundary layer equations.
The UPMWAKE model \cite{crespo1988} is a Navier-Stokes code for the far wake based on a
turbulence model.
Magnusson \cite{magnusson1999} uses momentum theory and blade element theory,
along with Prandtl's approximation, to provide a numerical model for the near-wake.
Some of the wake models mentioned have been integrated into wind farm design software.

This work is motivated by the desire to produce a model that more accurately predicts the near-wake region.
The Jensen model provides a reasonable representation of the wake for the mid- and far-wake regimes,
but there is a clear discrepancy in the near-wake, with the Jensen model predicting an unphysical
drop-off in the centreline wake wind velocity, see for example references \cite{okulov2015}, \cite{mirocha2015}, \cite{gallacher2014}.
It is well known that the transverse velocity deficit profile can be represented by a single-Gaussian
function for the mid- and far- wakes, but that in the near-wake the profile resembles a double-Gaussian
function, with local minima at about 75$\%$ blade span \cite{vermeera2003}, \cite{aitken2014}.
Thus, it is reasonable to consider a double-Gaussian function as a candidate for the transverse
velocity deficit profile. However, this in itself is not sufficient to generate physically realistic solutions.
These are obtained by subsequent, appropriate mixing of a wake velocity deficit solution and the freestream velocity.
It is natural that such an adjustment should be required as only a fraction of the wind flow fluid particles
passing through the rotor disk are affected by the blades \cite{burton2011}, and the amount by which they
are affected depends upon the pitch angle of the rotor blades.
The double-Gaussian velocity deficit and the mixing of solutions are not part of the basic actuator disk scenario. Consideration is given only to wind turbine wakes in
zero yaw situations. The predictions of this simple model are compared to real-scale wind turbine
wake measurements.

According to blade element momentum theory, the axial flow induction factor varies radially and azimuthally
across the rotor disk \cite{burton2011}. Here, for a particular wind flow regime, and in accordance with the actuator disk theory,
a {\it single} value of axial flow induction factor shall be taken to be representative of the wind flow at the rotor disk.
The proposed wake model is {\it axially symmetric} about the rotor axis. Inclusion of wind shear will
be dealt with in a subsequent work.

The new wake model is derived in section \ref{sec:wakemodelderivation}. In section \ref{sec:comparison}
the model is compared to wake measurements and estimates of quality of the fit are made.
A comparison is also made with the Jensen model. An overview of the results is given in section
\ref{sec:conclusions}.

\section{Wake model derivation}
\label{sec:wakemodelderivation}
The technique presented in Bastankhah and Port\'{e}-Agel \cite{bastankhah2014} shall be followed closely.

\subsection{Velocity deficit}
Consideration is given only to wakes in zero yaw situations:
All velocities are assumed to be in the axial direction.
In the following $U_\infty$ shall denote the incoming wind velocity
(the freestream velocity), $U$ is the wake velocity in the streamwise direction, and $U_D$ the velocity at the rotor disk. The normalized velocity deficit is defined as
\be
{\Delta U \over U_\infty} = {(U_\infty - U) \over U_\infty}.
\ee
The axial flow induction factor is defined as \cite{burton2011}
\be
a = 1 - {U_D \over U_\infty}.
\ee
The corresponding (theoretical) thrust coefficient for a wind turbine is defined as \cite{burton2011}
\be
C_T = 4a(1-a)
\ee
although an alternative empirical relationship between $C_T$ and $a$ can also be assumed, as described in chapter 4 of \cite{burton2011}.

The velocity deficit is taken to be of the form
\be
{\Delta U \over U_\infty} = C(x) f(r, \sigma(x))
\ee
where $C(x)$ is an arbitrary function of downwind distance $x$, and $f$ is an arbitrary function of the radial distance from the rotor axis $r$ and $\sigma$, the wake cross-section at downwind distance $x$.

The double-Gaussian profile is given by
\be
f(r, \sigma(x)) = \case{1}{2} [\exp D_+ + \exp D_-],
\qquad
D_\pm = - \case{1}{2} \sigma^{-2}(x) (r \pm r_0)^2
\label{eq:double-gaussian}
\ee
where $r_0$ is the radius corresponding to the locii of the Gaussian minima and
$\sigma(x)$ is a cross-section function.
As expected, in the case where $r_0 = 0$, the above coincides with (12) in \cite{bastankhah2014}, i.e.,
\be
f(r, \sigma(x)) = \exp [- \case{1}{2} \sigma^{-2}(x) r^2].
\ee

\subsection{Momentum conservation}
The equation of mean momentum flux through the rotor disk plane
(Equation (4.1.26) of Tennekes and Lumley \cite{tennekes1972}) is
\be
\rho \int U (U_\infty - U) dA = T
\label{eq:mass-and-momentum-conservation}
\ee
where $\rho$ is the density of the air, $A$ is the cross-sectional area in the plane of the rotor blade motion, and $T$ is the total force over the wind turbine, given by
\be
T = \frac{1}{2} C_T \rho A_0 U^2_\infty.
\ee
The quantity $C_T$ is the thrust coefficient defined above,
and $A_0$ is the area swept out by the turbine rotors, i.e.,
$A_0 = \pi d^2_0 / 4$ where $d_0$ is the wind turbine rotor diameter.

Following a similar procedure to that given in \cite{bastankhah2014},
equation (\ref{eq:mass-and-momentum-conservation}) is integrated.
The integration is performed over the entire rotor disk plane, i.e.,
a disk of {\it infinite} radius $r$. The area $A = \pi r^2$ and it follows that
\be
dA = 2 \pi r dr.
\ee
Thus, in order to evaluate equation (\ref{eq:mass-and-momentum-conservation}), one must calculate the value of the integral
\be
I = \int^{r = \infty}_{r = 0} U (U_\infty - U) dA.
\ee
The ansatz for the velocity deficit will now be used to obtain
solutions of equation (\ref{eq:mass-and-momentum-conservation}).

\subsection{Integration of equation of mean momentum flux}
The explicit expression for this integral is
\be
I = \pi \int^{r = \infty}_{r = 0} U^2_\infty C(x)
\biggl( \exp D_+ + \exp D_- - \case{1}{2} C(x) H r dr \biggr)
\ee
where
\be
H = \exp (2D_+) + \exp (2D_-) + 2 \exp (D_+ + D_-).
\ee
Upon integration this becomes (see Appendix)
\be
I = \pi U^2_\infty C(x) [ M - \case{1}{2} C(x) N ]
\ee
where
\begin{align}
M &= 2 \sigma^2 \exp(-\case{1}{2} \tau^2) + \sqrt{2 \pi} r_0 \sigma [ \erfc(\tau / \sqrt{2}) -1]
\nonumber \\
N &= \sigma^2 \exp(-\tau^2) + \case{1}{2} \sqrt{\pi} r_0 \sigma [ \erfc(\tau) -1]
\nonumber \\
\tau &= r_0 \sigma^{-1}. \label{eq:MN-definition}
\end{align}
Hence, equation (\ref{eq:mass-and-momentum-conservation}) is
\be
C^2(x) N - C(x)M + \case{1}{8} C_T d^2_0 = 0
\ee
whose solution is
\be
C_\pm(x) = {M \pm \sqrt{M^2 - \case{1}{2} N C_T d^2_0} \over 2N}
\label{eq:quadratic-solution}
\ee
The solution of equation (\ref{eq:mass-and-momentum-conservation}) is given in terms of a Gaussian-type velocity deficit, with cross-section $\sigma$ chosen as
\be
\sigma = k^* x^n + \epsilon
\label{eq:cross-section}
\ee
where $k^*$ and $\epsilon$ are constants and $n$ is an appropriately chosen exponent,
see \cite{frandsen2006}, \cite{bastankhah2014}, \cite{tennekes1972}.
Classical theories of shear flows predict $n = 1/3$ \cite{tennekes1972} and this value
shall be adopted here.
Note that $\sigma$ is the cross-section width of the single-Gaussian, rather than the cross-section
of the full double-Gaussian.
The velocity solutions $U_\pm$ of equation (\ref{eq:mass-and-momentum-conservation}) are
\be
U_\pm = U_\infty \biggl(1 - C_\pm(x) f(r, \sigma(x)) \biggr).
\ee

\begin{figure}[h!]
  \centering
  \includegraphics[width=1.0\textwidth]{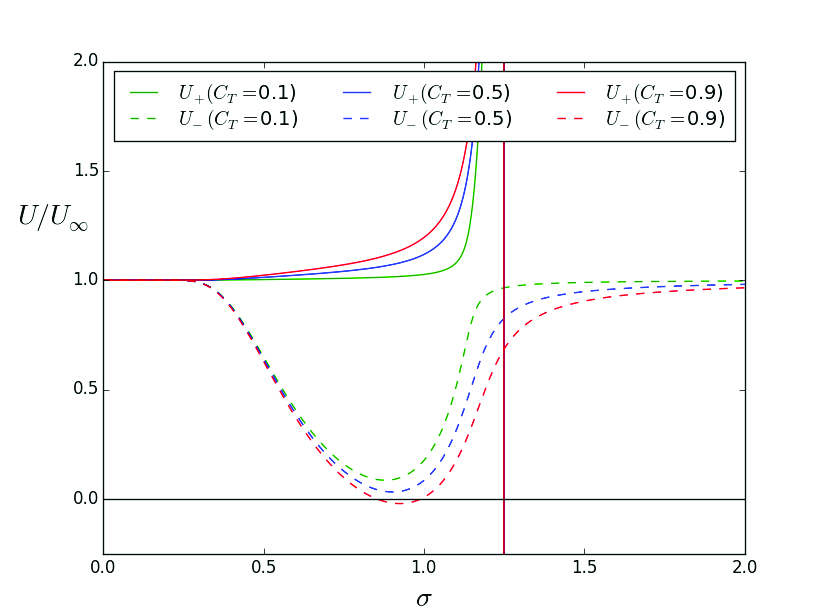}
  \caption{Wake centreline hub height horizontal normalized velocity profiles for the mean momentum
  flux equation solutions $U_+$ and $U_-$, for a selection of values of thrust coefficient $C_T$.
  The abscissa is the cross-section $\sigma$.
  The $U_+$ and $U_-$ solutions are represented by solid and dashed lines respectively.
  The $U_-$ solutions have local minima with values of approximately 0.1, 0.05 and -0.05 for $C_T$
  values $0.1$, $0.5$ and $0.9$ respectively.}
  \label{fig:diagram3}
\end{figure}

One can carry out analysis in terms of the new dimensionless variable
\be
X = \sigma / r_0.
\ee
In that case one can define
\ba
\begin{aligned}
{\cal M} &= M/r^2_0 = X \biggl( 2 X \exp(-\case{1}{2} X^{-2}) + \sqrt{2 \pi} [ \erfc(X^{-1} / \sqrt{2}) -1] \biggr)
\nonumber\\
{\cal N} &= N/r^2_0 = X \biggl( X \exp(-X^{-2}) + \case{1}{2} \sqrt{\pi} (\erfc(X^{-1})-1) \biggr)
\end{aligned}
\ea
and the solutions (\ref{eq:quadratic-solution}) can be expressed as
\be
{\cal C}_\pm(X) = {{\cal M} \pm \sqrt{{\cal M}^2 - \case{1}{2} {\cal N} C_T d^2_0 r^{-2}_0 } \over 2 {\cal N}}.
\ee
The corresponding velocity solutions $U_\pm (X)$ are plotted for representative values of the
thrust coefficient $C_T$ in Figure \ref{fig:diagram3}. Here and throughout, $r_0$ is taken
to be 75$\%$ blade span in accordance with the empirical measurements in references
\cite{vermeera2003} and \cite{aitken2014}, i.e., $r_0 = 0.75 d_0/2$.
The following facts are noted,
\begin{figure}[t!]
  \centering
  \includegraphics[width=0.75\textwidth]{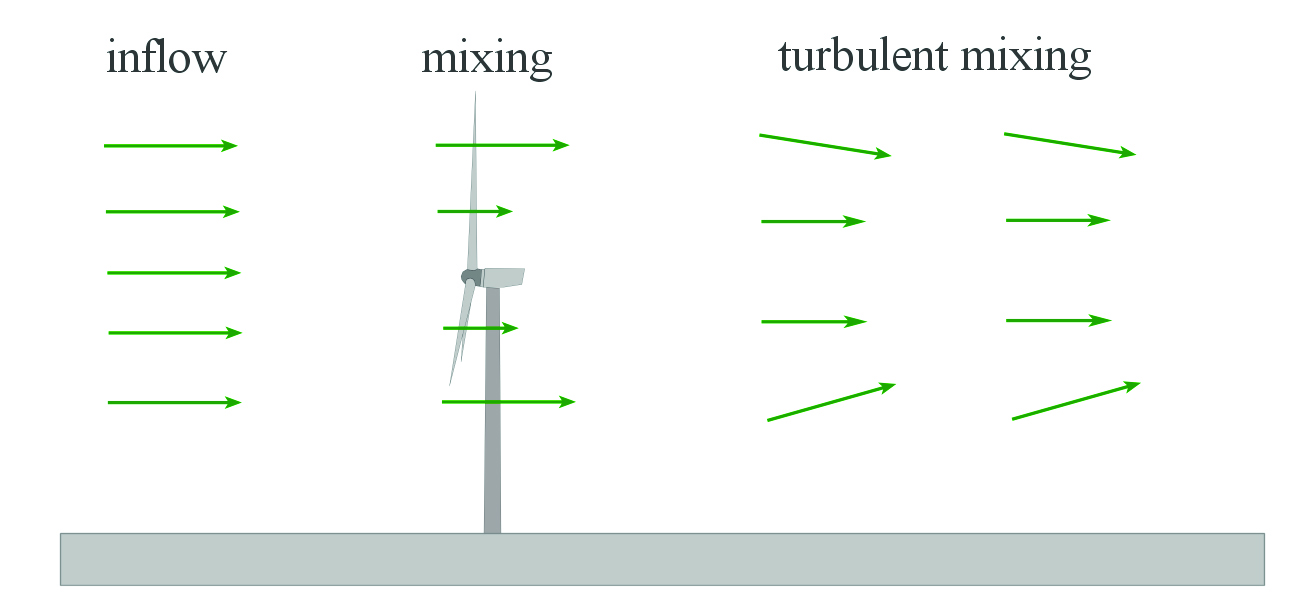}
  \caption{The mixing at the rotor and near wake of the wind turbine is governed by
  equation (\ref{eq:mixing-solution}) and is distinct from the lateral turbulent mixing that occurs
  further downstream.}
  \label{fig:diagram4}
\end{figure}
\begin{enumerate}[i]

    \item The $U_+$ solutions are unphysical.

    \item The $U_-$ solutions show a clear local minimum for all non-zero values of $C_T$.

    \item The $U_-$ solutions have the desired topological features, but the values of wake velocity at the local minima of the $U_-$ solutions are far lower than those obtained from empirical measurements.

    \item The ambient background velocity (freestream velocity) $U_\infty$ is a (trivial) solution
    of the equation of mean momentum flux (\ref{eq:mass-and-momentum-conservation}) for $T = 0$.

\end{enumerate}

The key to reconciling the above theory with the true measured wake velocity profiles is to consider a linear combination of solutions (mixing).

\subsection{Velocity deficit solution: Mixing in the wake}
Motivated by the facts i - iv noted above, it is conjectured that the real velocity solution in the wake
is a linear combination of the solution $U_-$ and
the freestream velocity $U_\infty$, as depicted in Figure \ref{fig:diagram4}.
This real solution is
\be
U = c_- U_- + c_\infty U_\infty, \qquad c_- + c_\infty = 1
\label{eq:mixing-solution}
\ee
for constants $c_-$ and $c_\infty$. Both solutions $U_-$ and $U_\infty$ are present in the wake,
the combination of the two states through mixing gives rise to the final wind velocity profile.
It is noted that a $c_+ U_+$ term is {\it not} included because the resultant solutions $U$ will
always be unphysical. The following points are noted.

\begin{enumerate}[i]

    \item The equation of mean momentum flux together with the actuator disk theory provide one component of the solution ($U_-$). It is conjectured that this represents the component actually affected by the rotor blades.

    \item The other component is provided by the trivial solution, i.e. the solution had there been no wind turbine, namely that corresponding to the ambient background velocity (freestream velocity) $U_\infty$.

\end{enumerate}
The constant $c_-$, and hence $c_\infty$, will vary with $C_T$. Recall that in the case of
high wind velocity ($U_\infty \gtrsim 12 m/s$) a relatively low proportion of energy is
being extracted from the wind since the blades are in pitch/stall mode. In that case $c_-$ will
be relatively small and $c_\infty$ relatively large. In the low $U_\infty$ regime $c_-$
will be relatively large.

As stated in section 3.8 of \cite{burton2011}, much of the theory and
analysis in the literature is based on the assumption that there is a sufficient number
of blades in the rotor disk for every fluid particle in the rotor disk to interact with a blade.
However, with a small number of blades most fluid particles will pass {\it between} the blades
and the loss of momentum by a fluid particle will depend upon its proximity to a blade.
Further, close to the blade tips, the tip vortex causes very high values of $a$ such that, locally,
the net flow past the blade is in the upstream direction. These facts are qualitatively consistent with
the proposed model and give credence to the method of construction of solutions above and their interpretation.

\begin{figure}[t!]
  \centering
  \includegraphics[width=0.8\textwidth]{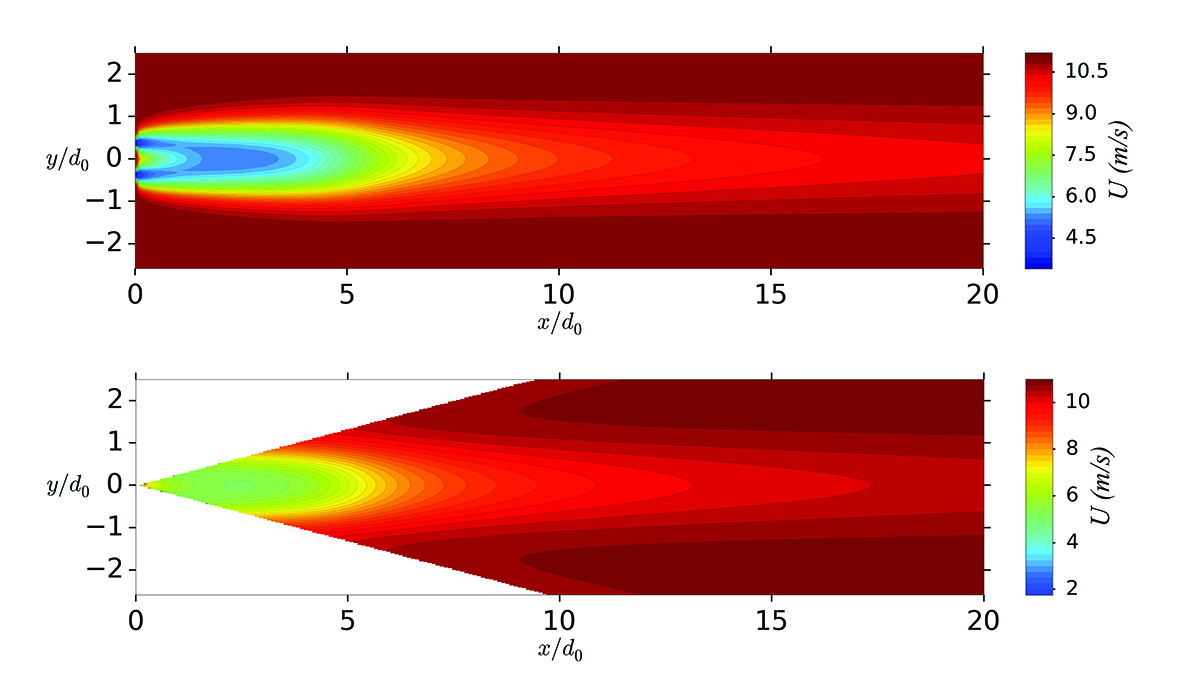}
  \caption{The top plot shows the wind velocity for the hub height horizontal
  section through the wake model. The lower plot shows the corresponding radial
  wind velocity, as would be measured by a nacelle-mounted LiDAR with $30^\circ$ horizontal scanning arc.
  $U_\infty = 11 ms^{-1}$.}
  \label{fig:diagram2}
\end{figure}

Figure \ref{fig:diagram2} shows the wind velocity for the hub height horizontal
section through the wake model, and the corresponding radial wind velocity, as would be
measured by a nacelle-mounted LiDAR.
Worthy of note is the ``jet structure" in the near-wake, with local
minima at about 75$\%$ blade span \cite{vermeera2003}, \cite{aitken2014}, arising from the
double-Gaussian wind velocity deficit distribution.
The model exhibits the characteristic wake expansion downwind of the turbine rotor.

\begin{figure}[t!]
  \centering
  \includegraphics[width=1.0\textwidth]{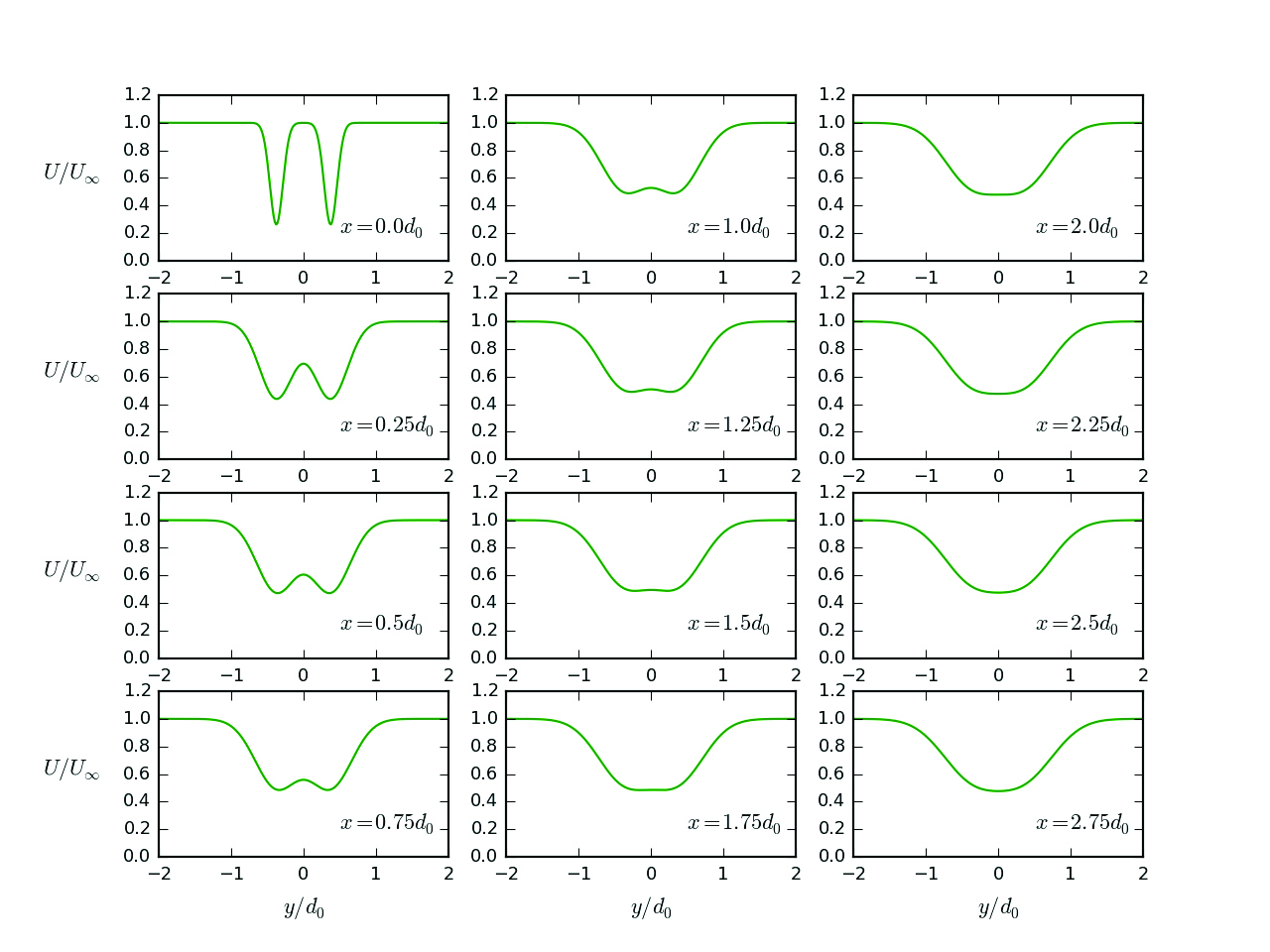}
  \caption{The wind velocity profiles for the hub height horizontal cross-sections through
  the wake model, for $U_\infty = 11 ms^{-1}$, for various downwind distances $x$.
  The relevant parameter values are given in Table \ref{tab:parameters}.
  It is clear that there is a transition from double-Gaussian to single-Gaussian distribution
  at a downwind distance of about $2.5d_0$.}
  \label{fig:diagram6}
\end{figure}

Figure \ref{fig:diagram6} shows the wind velocity profiles
for the hub height horizontal cross-sections through the wake model,
for $U_\infty = 11 ms^{-1}$, for various downwind distances.
It is clear that there is a transition from double-Gaussian to single-Gaussian distribution
at a downwind distance of about $2.5d_0$.

In summary, the real wake velocity solution is
\be
U = c_- U_\infty \biggl(1 - C_-(x) f(r, \sigma(x)) \biggr) + c_\infty U_\infty,
\qquad c_- + c_\infty = 1.
\ee
With a further little bit of algebra this can be written as
\be
U = U_\infty \biggl(1 - c_- C_-(x) f(r, \sigma(x)) \biggr).
\label{eq:mixing-solution-summary}
\ee
The functions $C_-(x)$, $f(r, \sigma(x))$ and $\sigma(x)$ are given by equations
(\ref{eq:quadratic-solution}), (\ref{eq:double-gaussian}) and (\ref{eq:cross-section}) respectively.
The function $C_-(x)$ is given in terms of the functions $M$ and $N$ defined
in equations (\ref{eq:MN-definition}).
The real wake velocity solution depends upon several parameters:
The wind turbine rotor diameter $d_0$, the radial location of the local minimum
which has been determined empirically as $r_0 = 0.75 d_0/2$,
$C_T$ which is fixed by the wind turbine's thrust characteristics and varies with inflow wind
speed $U_\infty$, and the parameters $k^*$, $\epsilon$ and $c_-$ are obtained from fitting
(see next section).

It can be seen from equation (\ref{eq:mixing-solution-summary}) that the method is mathematically
equivalent to a simple re-scaling of the value of $C_-(x)$ by the amount $c_-$.

\section{Comparison with wake data measurements}
\label{sec:comparison}
The wake model was compared to measured wind turbine wake data obtained in the analysis of
Gallacher and More \cite{gallacher2014}, which used nacelle-mounted LiDAR measurements to examine wake
length, width and height for a single wind turbine with $d_0 = 116$ m.

\begin{figure}[t!]
  \centering
  \includegraphics[width=0.85\textwidth]{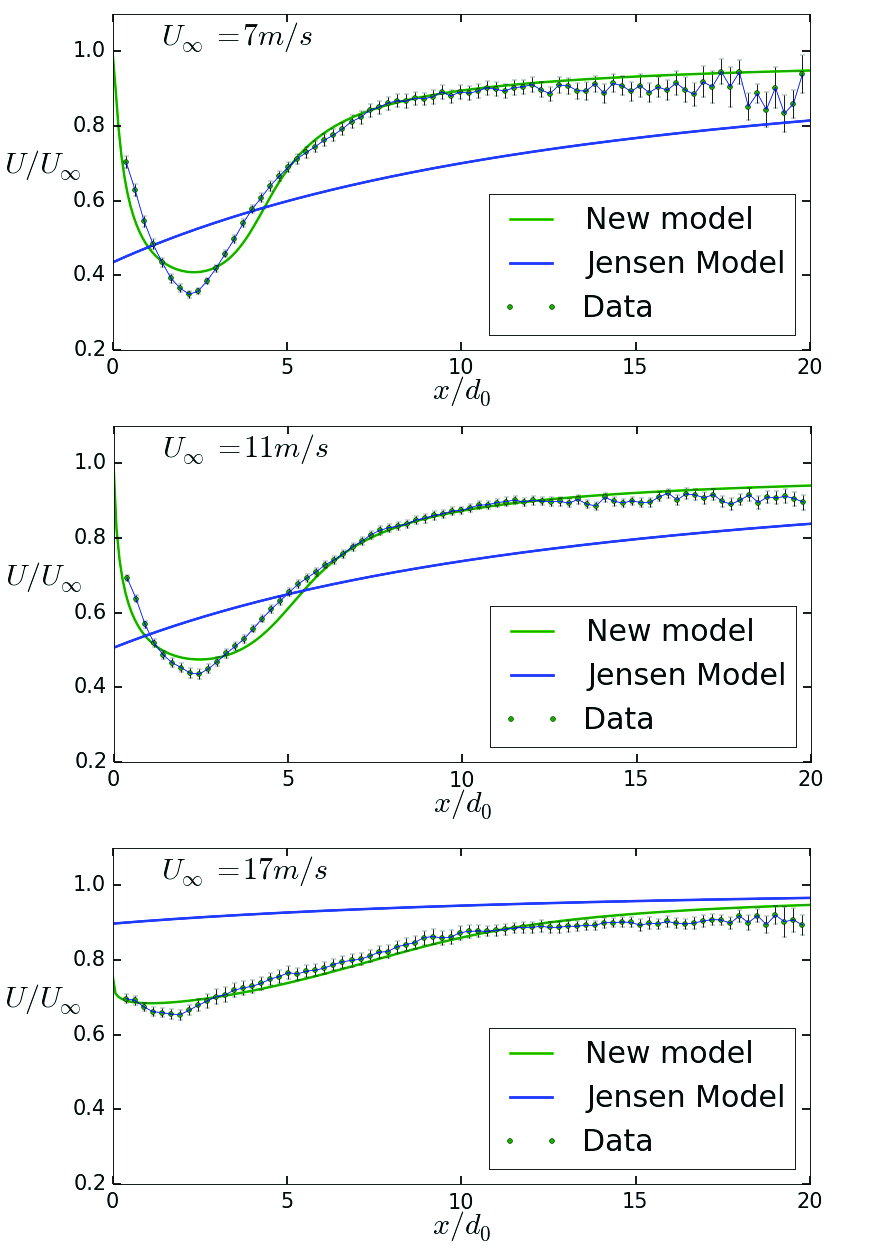}
  \caption{Horizontal normalized velocity profile for the wake centreline at hub height for the ($10$ minute averaged) wake data \cite{gallacher2014},
  the Jensen model (blue), and the newly proposed model (green).
  The velocity profiles are shown for $U_\infty = 7, 11$, and $17 ms^{-1}$.
  The rotor diameter $d_0 = 116$ m.
  Table \ref{tab:parameters} gives the corresponding best fit parameter values.}
  \label{fig:diagram1}
\end{figure}

Figure \ref{fig:diagram1} shows the wake centreline hub height velocity profile of the new
wake model for $U_\infty = 7, 11$, and $17 ms^{-1}$, in close agreement with the ($10$ minute averaged)
measured wake data. Table \ref{tab:parameters} gives the corresponding best
fit parameter values. The values of $C_T$ were looked up for each inflow windspeed based
on the wind turbine data.
The corresponding Jensen model (see, for example, equation (1) of \cite{bastankhah2014}),
\be
{\Delta U \over U_\infty} = \biggl(1 - \sqrt{ 1 - C_T} \biggr)
\biggl(1 + {{2 k_{wake} x} \over d_0} \biggr)^{-2}
\ee
with wake decay coefficient $k_{wake} = 0.038$ \cite{gallacher2014}, is shown for comparison.
(Alternative values are given in, for example, reference \cite{bastankhah2014}.)
The predicted velocity profile shows a clear local minimum at a distance of about $2$ rotor diameters,
in contrast to the Jensen model.

\begin{figure}[t!]
  \centering
  \includegraphics[width=0.9\textwidth]{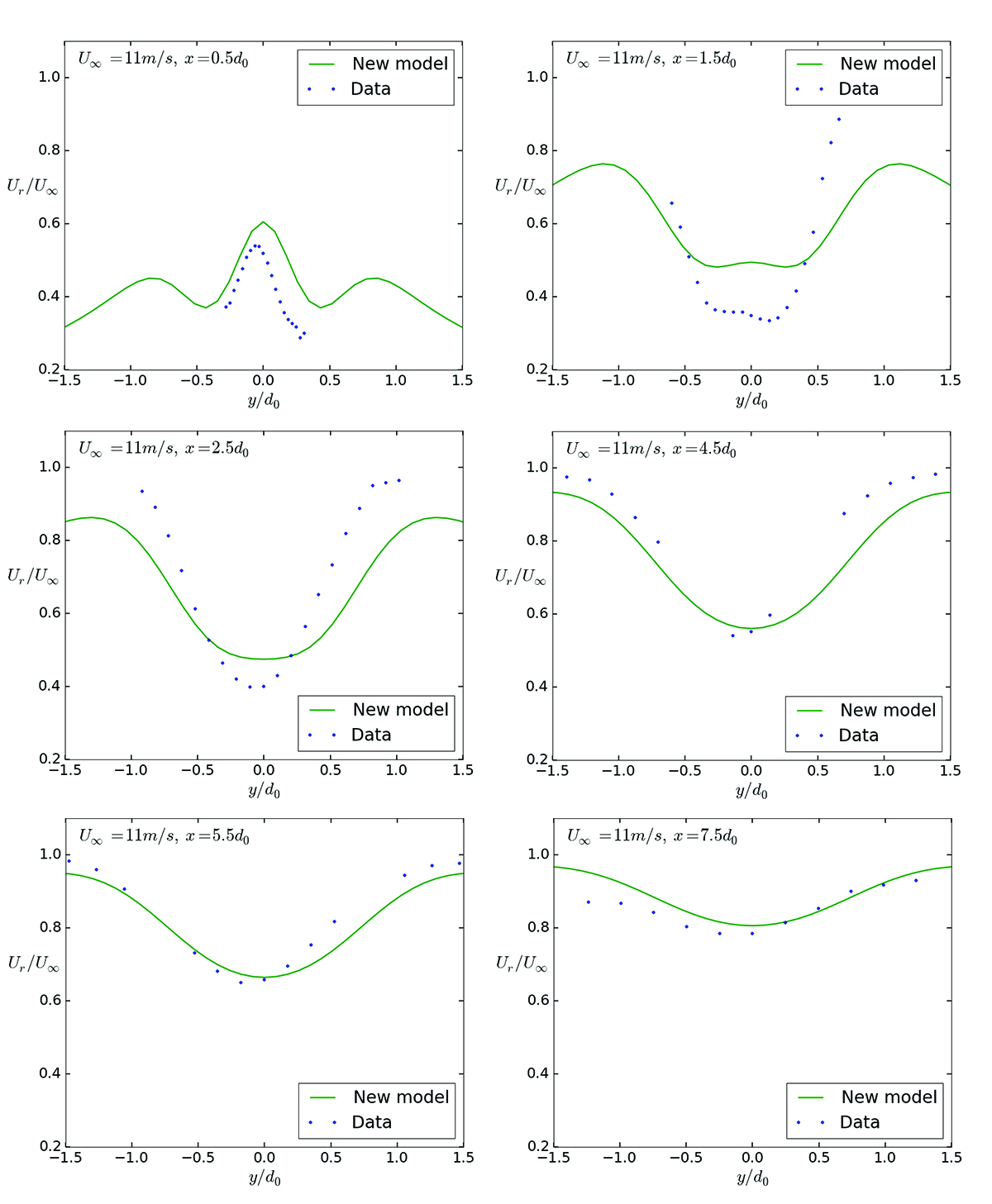}
  \caption{Horizontal normalized nacelle-mounted LiDAR {\it radial velocity}
  profile for the wake cross-sections at hub height for the ($10$ minute averaged)
  wake data \cite{gallacher2014}, and the newly proposed model, for $U_\infty = 11 ms^{-1}$ and selected downwind distances $x$,
  based on the {\it centreline hub height best fit} parameter values given in table \ref{tab:parameters}. The rotor diameter $d_0 = 116$ m.
  The magnitude of the LiDAR measured radial velocity tends to zero for large distances from the wake centreline,
  in contrast to the situation in Figure \ref{fig:diagram6}. This effect is particularly evident for small downwind distances
  and is due to the cosine factor arising as a result of the angular dependence of the LiDAR scan geometry.}
  \label{fig:diagram5}
\end{figure}

A comparison was made between the nacelle-mounted LiDAR radial velocity measurements of \cite{gallacher2014}
and the corresponding radial velocities given by the wake model.
Figure \ref{fig:diagram5} shows the horizontal normalized {\it radial velocity}
profiles for the wake cross-sections at hub height for the ($10$ minute averaged)
wake data, and the newly proposed model, based on the
{\it centreline hub height best fit} parameter values given in table \ref{tab:parameters}.

\begin{table}
\caption{\label{tab:parameters}
The model parameters, given to 3 significant figures, for the fitting to the
measured centreline hub height wake data. The rotor diameter $d_0 = 116$ m.}
\footnotesize\rm
\begin{tabular*}{\textwidth}{@{}l*{15}{@{\extracolsep{0pt plus12pt}}l}}
\br
Windspeed bin (m/s) & $U_\infty (m/s)$ & $a$ & $C_T$ & $k^*$ & $\epsilon$ & $c_-$ \\
\mr

6 - 8 & $7$ & $0.293$ & $0.828$ & $6.54$ & $0.040$ & $0.437$ \\

10 - 12 & $11$ & $0.256$ & $0.761$ & $5.40$ & $0.090$ & $0.400$ \\

16 - 18 & $17$ & $0.053$ & $0.201$ & $2.18$ & $0.250$ & $0.313$ \\

\br
\end{tabular*}
\end{table}

\section{Discussion}
\label{sec:discussion}
It can be seen from Figures \ref{fig:diagram1} and \ref{fig:diagram5} that the new proposed
wake model is in close agreement with the measured data.
In Figure \ref{fig:diagram1} the proposed model performs better than the corresponding
Jensen model which, when compared to the measured data, exhibits an unphysical drop
off in the centreline wake velocity.
Figure \ref{fig:diagram5} shows the horizontal hub height radial velocity profiles,
and those predicted by the wake model, based on the {\it centreline hub height best fit}
parameter values. The model performs reasonably well in the near wake region, exhibiting the
expected local minima, and showing better agreement with increasing downwind distance.
The deviations from perfect fits in the profiles in Figure \ref{fig:diagram1} could in part be
attributed to a slightly asymmetric wake profile about the rotor axis, evident in Figure \ref{fig:diagram5},
and the presence of other wind turbines.

The proposed wake model is limited by being axially symmetric about the rotor axis.
However, the inclusion of vertical wind shear shall be considered in a subsequent publication.
The comparison given above are based upon 1D fitting, and only for the wake centreline hub
height measurements. Future improvements can be made by simultaneous 2D fitting to
wake cross-section hub height data, and possibly full 3D fitting to an entire wake.

\section{Conclusions}
\label{sec:conclusions}
An analytical wind turbine wake model has been proposed to predict the wind velocity distribution
for all distances downwind of a wind turbine, including the near-wake.
The model is derived by applying conservation of mass and momentum in the context of actuator disk theory,
and assuming a distribution of the double-Gaussian type (\ref{eq:double-gaussian}) for the velocity deficit in the wake.
The physical solutions are obtained by appropriate mixing of the waked- and freestream velocity
deficit solutions through equation (\ref{eq:mixing-solution}), reflecting the fact that only a portion of the fluid particles passing
through the rotor disk will interact with a blade. As mentioned previously, this  is mathematically
equivalent to a simple re-scaling of the velocity deficit.
The wake model can be used by selecting a wind speed regime, and consulting Table \ref{tab:parameters}
to obtain the corresponding parameter values to provide an appropriate set of values
$\{ C_T, k^*, \epsilon, c_- \}$.

The proposed wake model is a development of actuator disk theory, but without the complications
of full blade element momentum theory.
It has potential for applications to windfarm modeling, with the possibility of incorporating
multiple wind turbine wakes.
With further development there is the potential for comparison with Large-eddy simulations.
The model may also serve to illuminate the physics of the near-wake.
Turbulence intensity and dynamic features of the wake important for load modeling have
not been considered here. This wake model will be developed
further in a subsequent publication, where consideration will be given to inclusion
of wind shear, power production and yaw effects.

\section*{Acknowledgments}
We thank Roy Spence, Graham More, Alan Mortimer, and Bianca Shulte for useful discussions.

\appendix
\section*{Appendix}
\setcounter{section}{1}

Define
\be
J = \int^{r = \infty}_{r = 0} \biggl[ E_+ + E_- -\case{1}{2} C(x) (F_+ + F_- + 2G) \biggr] r dr
\ee
where
\ba
E_\pm = \exp D_\pm, \qquad F_\pm = \exp (2 D_\pm), \qquad G = \exp (D_+ + D_-),
\ea
so that the original integral can be written
\be
I = \pi U^2_\infty C(x) J.
\ee

Now consider each term separately. The upper limit of the integration will be taken as $r = R$, which will later be set to $\infty$. Define
\be
J_1 = \int^{r = R}_{r = 0} E_+ r dr
\ee
and the new variable
\be
r' = r + r_0, \qquad dr' = dr.
\ee
Then
\begin{align}
J_1 &= \int^{r' = R + r_0}_{r' = r_0} \exp (- \case{1}{2} \sigma^{-2} r'^2 ) (r' - r_0) dr'
\nonumber\\
&= \int^{r' = R + r_0}_{r' = r_0} \exp (- \case{1}{2} \sigma^{-2} r'^2 ) r'dr'
- r_0 \int^{r' = R + r_0}_{r' = r_0} \exp (- \case{1}{2} \sigma^{-2} r'^2 ) dr'.
\end{align}
These last integrals are of the form
\be
\int \exp (x^2) x dx \qquad \mbox{and} \qquad \int \exp (x^2) dx
\ee
respectively. The former is straightforward to integrate, and the second requires use of the complementary error function \cite{stephenson1990}
\be
\erfc(x) = {2 \over \sqrt{\pi}} \int^{t = \infty}_{t = x} \exp (-t^2) dt.
\ee
Making the change of variables $r' = \sqrt{2} \sigma t$ this becomes
\be
\erfc(x) = \sqrt{2 \over \pi} \sigma^{-1} \int^{r' = \infty}_{r' = r_0} \exp (-\case{1}{2} \sigma^{-2} r'^2) dr'
\ee
where $r_0 = \sqrt{2} \sigma x$.
Similarly, for
\be
J_2 = \int^{r = R}_{r = 0} E_- r dr
\ee
define the new variable
\be
r'' = r - r_0, \qquad dr'' = dr.
\ee
and so
\be
\fl J_2 = \int^{r'' = R - r_0}_{r'' = -r_0} \exp (- \case{1}{2} \sigma^{-2} r''^2 ) r''dr''
+ r_0 \int^{r'' = R - r_0}_{r'' = -r_0} \exp (- \case{1}{2} \sigma^{-2} r''^2 ) dr''.
\ee
Taking $R = \infty$, the sum of the first terms in $J_1 + J_2$
is
\be
2 \sigma^2 \exp (-\case{1}{2} \sigma^{-2} r_0^2).
\ee
It follows that the sum of the second terms in the integrals is equal to
\be
\sqrt{\pi \over 2} r_0 \sigma
\biggl[ \erfc \biggl({r_0 \sigma^{-1} \over \sqrt{2}} \biggr) - \erfc \biggl({- r_0 \sigma^{-1} \over \sqrt{2}} \biggr)
\biggr].
\ee
Further, since
\be
\erfc(-x) = 2 - \erfc(x)
\ee
it follows that this sum is equal to
\be
\sqrt{2\pi} r_0 \sigma
\biggl[ \erfc \biggl({r_0 \sigma^{-1} \over \sqrt{2}} \biggr) - 1 \biggr].
\ee
Thus
\be
J_1 + J_2 = 2 \sigma^2 \exp (-\case{1}{2} \sigma^{-2} r_0^2) + \sqrt{2\pi} r_0 \sigma
\biggl[ \erfc \biggl({r_0 \sigma^{-1} \over \sqrt{2}} \biggr) - 1\biggr].
\ee

Similarly, defining
\be
J_3 = \int^{r = \infty}_{r = 0} F_+ r dr, \qquad J_4 = \int^{r = \infty}_{r = 0} F_- r dr
\ee
it can be shown that
\be
J_3 + J_4 = \sigma^2 \exp (-\sigma^{-2} r_0^2) + \sqrt{\pi} r_0 \sigma
\biggl[ \erfc (r_0 \sigma^{-1}) - 1 \biggr].
\ee

Finally, defining
\be
J_5 = \int^{r = \infty}_{r = 0} G r dr
\ee
it follows that
\be
J_5 = \int^{r = \infty}_{r = 0} \exp [-\sigma^{-2} (r^2 + r_0^2)] r dr
= \case{1}{2} \sigma^2 \exp [-\sigma^{-2} r_0^2].
\ee

\end{document}